# Nonlinear quantum processes in superconducting resonators terminated by neon-focused-ion-beam-fabricated superconducting nanowires

J. A. Potter, O. W. Kennedy, J. C. Fenton and P. A. Warburton

London Centre for Nanotechnology,
University College London,
London, United Kingdom
p.warburton@ucl.ac.uk

*Abstract*—We have used a neon focused-ion-beam to fabricate both nanoscale Nb Dayem bridges and NbN phase-slip nanowires located at the short-circuited end of quarter-wavelength coplanar waveguide resonators. The Dayem bridge devices show flux-tunability and intrinsic quality factor exceeding 10,000 at 300 mK up to local fields of at least 60 mT. The NbN nanowires show signatures of incoherent quantum tunnelling of flux at 300 mK.

## I. Introduction

Superconducting nanowires with cross-sectional dimensions comparable to or less than the coherence length display nonlinear transport characteristics which can be exploited in quantum and classical electronics applications. The Dayem bridge [1] consists of a short region of suppressed superconducting order between two fully superconducting electrodes. Josephson coupling between the electrodes results in it behaving as a nonlinear inductor. If the superconducting nanowire has a high degree of structural disorder then the physics may be dominated by quantum phase-slips (QPS) [2,3]. The coherent-QPS nanowire is the charge-flux dual of the Josephson junction and it behaves as a nonlinear capacitor.

For quantum device applications it is critical that dissipative loss mechanisms at non-zero frequencies (which, for example, suppress the $T_1$ coherence lifetime) are minimised. The technological challenge is therefore to fabricate superconducting nanowires with widths a few tens of nanometres without significantly increasing the dissipative loss rate above that of the unpatterned film. Focused-ion-beams (FIB) are widely used for nanofabrication of superconductors [4,5], and traditionally made use of gallium ions from a liquid metal ion source. However, Ga-based FIB milling has undesirable effects such as gallium poisoning —giving degraded superconducting material and increased dissipation [6]— and gallium's high mobility —leading to long-timescale device variability [7]. More recently the gas-field ion source (GFIS) has been used to generate both helium and neon ion-beams, with the latter having sufficient mass to perform FIB milling of metallic thin films.

Here we demonstrate the use of neon-FIB milling for fabrication of superconducting nanowires. We have created nanoscale superconducting loops containing either (a) Dayem bridges in Nb films or (b) phase-slip nanowires (operating here in the incoherent regime) in disordered NbN films. These nanowires are embedded within quarter-wavelength superconducting coplanar waveguide (CPW) resonators so measurements of the resonant frequency and quality factor enable, respectively, the nonlinear properties of and losses in the nanowires to be extracted.

## II. Nanowire-terminated transmission lines

The impedance $Z_{in}$ seen at the input to a transmission line of length $l$ which is terminated by a load impedance $Z_L$ that is not equal to the characteristic impedance $Z_0$ of the line is given by

$$Z_{in} = Z_0 \frac{Z_L + iZ_0 \tan(2\pi v l/c)}{Z_0 + iZ_L \tan(2\pi v l/c)}, \qquad (1)$$

where $v$ is the frequency and $c$ is the phase velocity. At resonance the imaginary part of $Z_{in}$ is zero. Hence, modulation of the load inductance $Z_L$ results in tuning of the resonant frequency of the transmission line.

A DC-SQUID has flux-tunable inductance given by

$$L_k = \frac{\Phi_0}{4\pi I_c \left| \cos\left(\frac{\pi \Phi}{\Phi_0}\right) \right|}, \qquad (2)$$

where $I_c$ is the SQUID critical current at zero flux, $\Phi$ is the applied flux and $\Phi_0$ is the flux quantum. In the case of a

This work is supported by EPSRC (grant references EP/H005544/1, EP/K024701/1, EP/L015242/1 and EP/P510270/1).

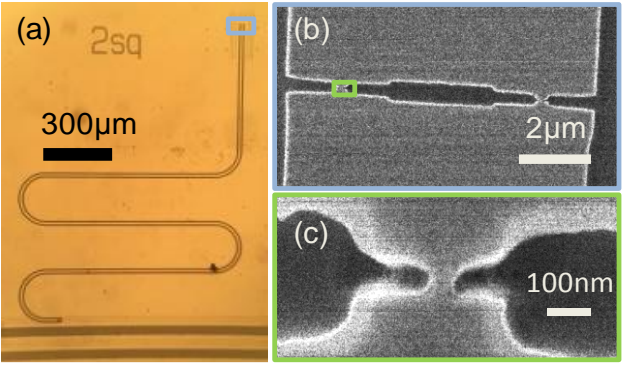

Fig. 1. Nb Dayem-bridge device. (a) Optical micrograph of the CPW meander resonator with feedline at the bottom of the image; (b) Helium FIB image of the highlighted region in (a) showing the SQUID loop; (c) Helium FIB image of the highlighted region in (b) showing a single Dayem bridge. In all images the Nb is bright and the exposed Si substrate is dark. Further details are given in [10].

SQUID-terminated transmission line the load impedance $Z_L$ is $i2\pi \nu L_k$.

If the Josephson junctions of the SQUID are replaced by two parallel superconducting nanowires then the form of the flux-modulation of the total inductive load is modified. In the case of a highly disordered superconductor such as niobium nitride there is a large nonlinear kinetic inductance which varies with DC current, $I$, as $L_k(I) = L_k(0)[1+I^2/I_*^2]$ [8]. Therefore the total kinetic inductance of the combined loop-nanowire system is given by

$$L_k = L_{k1}\left(1+\frac{I^2}{I_{*1}^2}\right) + L_{k2}\left(1+\frac{I^2}{I_{*2}^2}\right), \quad (3)$$

where $I_{*1}$ and $I_{*2}$ are the current scales which characterise the nonlinearity of the nanowires and the wider part of the loop respectively. These current scales are related to the transport critical current and so are related to each other by a geometric factor. Since $I = \Phi / L_k$, the flux-dependence of the total kinetic inductance can be calculated.

### III. NEON FOCUSED-ION-BEAM

A GFIS source [9] consists of a metallic (typically tungsten) tip which is cooled to below 100 K and sharpened by field emission to atomic dimensions. He or Ne gas is admitted to a pressure of ~$10^{-6}$ mbar. The potential applied to the tip is adjusted so that the electric field is only sufficiently large to ionise the inert gas atoms at the apex of the tip. This results in an inert gas beam of very high brightness which emerges from the three-atom tip, one atom of which is selected as an atomic ion source. Ions are then accelerated and focused onto the sample using conventional ion-beam optics. Typical neon beams in commercially available FIB systems have energy up to 30 kV, current up to 10 pA and beam diameter at the sample as small as 2nm. The beam energy, which is comparable to that in Ga-FIB, is controlled electrostatically. Its optimal value is chosen to give a balance between sputter yield and sample damage. The maximum beam current achievable with a neon beam is more than three orders of magnitude lower than that of commercially available gallium FIB systems, leading to significantly reduced volume milling rates for neon by comparison with gallium.

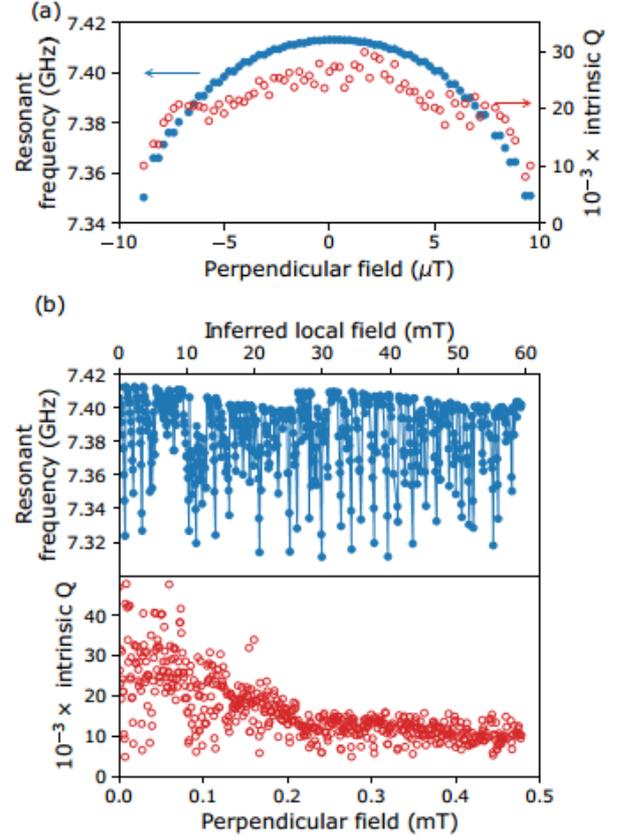

Fig. 2. (a) Dependence of the resonant frequency (blue, left axis) and intrinsic quality factor (red, right axis) of the Nb Dayem-bridge device on a magnetic field applied perpendicular to the SQUID loop. (b) The same measurements extended to higher field. The lower axis shows the applied field and the upper axis shows the inferred local field at the SQUID. The line is a guide to the eye. Data presented and analysed further in [10].

### IV. EXPERIMENTAL DETAILS

Films were grown by d.c. magnetron sputtering from a Nb target onto silicon (in the case of Nb) or *c*-axis oriented sapphire (in the case of NbN) substrates. For Nb deposition the sputter atmosphere is 100% argon, whereas for NbN deposition it is 50% argon − 50% nitrogen. The film thickness is ~50 nm for the Nb films and ~10 nm for NbN; the smaller thickness for NbN was chosen because the phase fluctuations leading to quantum phase-slip are enhanced in thinner films.

The larger-scale features of our devices were fabricated by electron-beam lithography and reactive ion etching in $SF_6$. At this stage an r.f. feedline and quarter-wavelength CPW resonators are defined. Multiple resonators with different resonant frequency are capacitively coupled to the same feedline, allowing multiplexed readout of up to 6 devices on the same chip. Each resonator is fabricated with a loop at the short-circuited end of the resonator, containing precursor 'microwires' with dimensions of a few hundred nanometres.

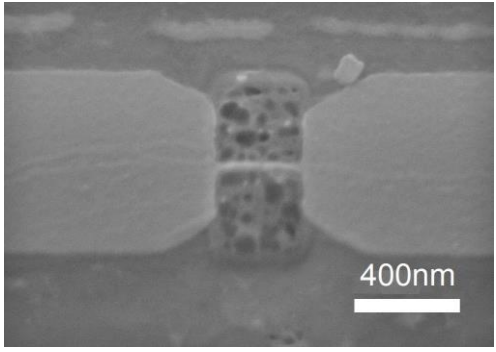

Fig. 3. Scanning electron microscope image of a NbN phase-slip nanowire. NbN appears brighter in contrast than the sapphire substrate.

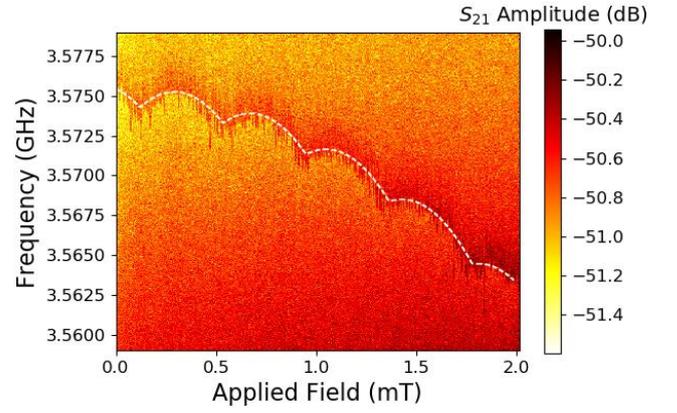

Fig. 4. Dependence of the transmission of the nanowire phase-slip device upon frequency and magnetic field applied perpendicular to the plane of the superconducting loop. The white dashed line shows the fit to the resonant frequency using (2) and (3), with $B^* = 14$ mT and $F = 2.4$.

Nanowires were fabricated by FIB milling the precursor microwires to the desired width using a 15-kV Ne beam with a current of ~2 pA. A dose of ≈2 nCµm$^{-2}$ was sufficient to clear the 50-nm Nb film, while a dose of 1 nCµm$^{-2}$ was used for the NbN devices. Ion milling on resistive or insulating substrates results in the build-up of positive charge on devices. This charge repels the incoming beam and can limit resolution. Charging effects during neon milling were mitigated by connecting the sample to a copper electrode via aluminium wire bonds. This was found to be particularly important for the NbN samples due to the high normal-state resistivity of niobium nitride thin films together with the insulating sapphire substrate. The sputter yield using the Ne beam was sufficient to fabricate nanowires in less than a minute; the short time was important to avoid charging-induced drift.

Measurements were performed in a $^3$He cryostat at a temperature $T \approx 300$ mK using an r.f. vector network analyser. Magnetic field was applied using a superconducting solenoid and a precision current source.

## V. NIOBIUM DAYEM BRIDGES

In Fig. 1 we show a Nb Dayem-bridge SQUID loop coupled to a λ/4 resonator. The width of the two bridges is around 50 nm. The magnetic-field dependence of the resonant frequency and quality factor of this structure at T = 300 mK are shown in Fig. 2(a). In an applied field of 10 µT, the resonant frequency is tuned by ~60 MHz from its zero-field value of 7.41 GHz. The intrinsic $Q$ of the resonance has a zero-field value of around 25,000. Another similar device fabricated on the same wafer has an intrinsic $Q$ of 120,000 [10]. To our knowledge this is the highest reported $Q$ of a SQUID-tunable resonator. Comparison with other on-chip resonators which do not include SQUIDs [10] confirms that the $Q$ in these devices is limited by losses in the "bulk" superconducting films, and that the additional losses due to presence of the nanowires and FIB milling are negligible by comparison.

In Fig. 2(b), we show the field dependence of the resonant frequency and $Q$ of the same device at higher applied fields. Jumps in the resonant frequency occur as additional flux quanta tunnel into the SQUID loop. The field periodicity of hysteretic SQUIDs together with the known area of the SQUID loop allow us to determine an estimate of the flux focusing from the diamagnetic superconducting ground planes, $F = 124$, and the field of maximal detuning in units of $\Phi_0$ allows us to extract a lower bound for the normalized inductance, $\beta_L = 2L_kI_c/\Phi_0 > 3.4$. The flux focusing factor allows us to rescale the field axis in Fig. 2(b) to give the local flux density, $B_{local} = FB_{applied}$. At local fields higher than 30 mT, the device not only retains its flux-tunability but also the $Q$ is flux-independent (within the noise level) and exceeds 10,000.

## VI. NIOBIUM NITRIDE PHASE-SLIP NANOWIRES

Fig. 3 shows a typical device used to investigate nanowire phase-slip physics in NbN. The nanowire shown is 350 nm long and 30 nm wide. A particularly striking feature of the image is the 'bubbled' area of substrate around the nanowire. This bubbling is caused by neon implanting in the sapphire substrate, disrupting the crystal structure and causing local swelling. Previous studies [11,12] into ion-beam-induced substrate bubbling have found the dimensions of these sub-surface bubbles to be beam-energy- and ion-dose- dependent. Our SEM images (Fig. 3) show bubbles with diameters of tens of nanometres, which is consistent with the previous findings [11,12]. It was also previously observed that the bubbles stand a few tens of nanometres above the substrate surface, and in our case this height is greater than the thickness of the NbN film. In what follows, we report measurements of this device at $T = 300$ mK.

The critical temperature of the NbN was measured to be 8.5 K. Our measurements therefore are made at temperatures far below the predicted value of the crossover temperature below which the rate of quantum activation of phase slips exceeds the thermal rate [13]. The intrinsic $Q$ for this device is 2,000 whereas, prior to the neon FIB milling stage, it was 13,000. Indeed, it has previously been shown that neon irradiation can have a damaging effect on resonator quality factor [14]. The higher loss in the NbN devices by comparison with the Nb devices discussed above, even before Ne milling, may be associated with the high level of structural disorder and low film thickness [15], both of which are necessary to maximise the QPS rate. Nevertheless, the $Q$ of the neon-milled devices is two orders of magnitude larger than previously reported coherent-QPS devices based on indium oxide films [16] and somewhat larger than other NbN CQPS devices [17].

The magnetic-field dependence of the spectral response of this device is shown in Fig. 4. We observe two effects: (i) a slow decrease of the resonant frequency as the field increases. This matches the expected parabolic field-dependence of the kinetic inductance of the "bulk" NbN film, which (unlike for the Nb devices reported above) is significantly larger than the geometric inductance; (ii) a periodic modulation of the resonant frequency with period 0.42 mT, from which a flux focusing factor of $F = 2.4$ can be extracted. We argue that this periodic modulation arises from incoherent quantum phase-slip events occurring in the NbN nanowires, each allowing a single flux quantum to tunnel into the loop, and occurring when the flux difference between the inner and outer regions of the loop is equal to $\Phi_0/2$. We can rule out the presence of a weak-link Josephson junction at some point along the length of the nanowires, as this would lead to hysteretic tuning similar to that observed in our Nb SQUID devices.

The dashed line in Fig. 4 shows a fit to the data using (2) and (3), modified to include the parabolic field-dependence parametrised by a phenomenological field-scale $B^*$. The fitting parameters are the flux focusing factor and $B^*$. While we observe none of the avoided level crossings which would characterise coherent quantum phase-slip, the quality of the fit and the lack of hysteresis provide good evidence that incoherent quantum phase-slips are occurring in the nanowires. However, at this stage we cannot definitively rule out all other potential mechanisms for periodic flux tunnelling.

## VII. CONCLUSIONS

We used neon FIB to fabricate nanowires of width less than 50 nm which exhibit nonlinear quantum effects whilst maintaining low loss. The field resilience of the Nb Dayem bridges (shown here up to a perpendicular local field of 60 mT) suggests that they may be useful for application as a readout technology for spin qubits [10]. Furthermore, we have made measurements of incoherent flux tunnelling in NbN nanowires at 300 mK, which is a key milestone towards future studies of coherent quantum phase-slips in these devices. Future work will focus on minimising the losses at dilution-refrigerator temperatures where we expect CQPS phenomena to be observable.

ACKNOWLEDGMENT

The assistance of Jonathan Burnett, Nick Constantino, John Morton and Eva Dupont-Ferrier is gratefully acknowledged.